\documentclass[aps,pre,floatfix,twocolumn,superscriptaddress]{revtex4-1}

\usepackage[english]{babel}
\usepackage{amsmath}
\usepackage{amssymb,amsfonts,latexsym}
\usepackage[activate=normal]{pdfcprot}
\usepackage{hyphenat} 
\usepackage[final,hiresbb]{graphics}
\usepackage{color}
\usepackage[hidelinks]{hyperref}

\definecolor{blue}{rgb}{0,0,1}
\definecolor{red}{rgb}{1,0,0}
\definecolor{green}{rgb}{0,0.5,0}
\newcommand{\ms}[1]{\textcolor{black}{#1}}
\newcommand{\od}[1]{\textcolor{black}{#1}}

\newcommand{\bfn}{{\mathbf{\hat n}}}
\newcommand{\bfr}{{\mathbf{r}}}
\newcommand{\bfv}{{\mathbf{v}}}

\begin{document}
\title{Spontaneously flowing crystal of self-propelled particles}
\newcommand{\eccm}{EC2M, UMR Gulliver 7083 CNRS, ESPCI ParisTech, PSL Research University, 10 rue Vauquelin, 75005 Paris, France}
\newcommand{\pct}{PCT, UMR Gulliver 7083 CNRS, ESPCI ParisTech, PSL Research University, 10 rue Vauquelin, 75005 Paris, France}
\author{Guillaume \surname{Briand}}
\affiliation{\eccm}
\author{Michael \surname{Schindler}}
\affiliation{\pct}
\author{Olivier \surname{Dauchot}}
\affiliation{\eccm}

\date{\today}

\begin{abstract}
We experimentally and numerically study the structure and dynamics of a
mono-disperse packing of \od{spontaneously aligning} self-propelled hard disks.
The packings are such that their equilibrium counterparts form perfectly
ordered hexagonal structures. Experimentally, we first form a perfect crystal
in an hexagonal arena which respects the same crystalline symmetry. Frustration
of the hexagonal order, obtained by removing a few particles, leads to the
formation of a rapidly diffusing ``droplet''. Removing more particles, the
whole system spontaneously forms a macroscopic sheared flow, while conserving
an overall crystalline structure. \od{This flowing crystalline structure, which
we call a ``rheo-crystal'' is made possible by the condensation of shear along
localized stacking faults.} Numerical simulations very well reproduce the
experimental observations and \od{allow us to explore the parameter space}.
They demonstrate that the rheo-crystal is induced neither by frustration nor by
noise. They further show that larger systems flow faster while still remaining
ordered.
\end{abstract}

\maketitle
\ms{Crystals, translationally periodic structures, are known for their solid
behavior: they sustain their own shape. Liquids are disordered and flow to
adopt the shape of their container. Here, we use self-propelled hard disks to
show that this common knowledge is undermined in the context of active matter.}

\od{In the absence of alignment, the crystallization of self-propelled disks,
with repulsive interactions, follows the two-step scenario reported at
equilibrium~\cite{Bialke:2012cw,Klamser:2018vk}. The situation is far less
clear when alignment between particles overcomes rotational diffusion. In the
fluid phase, polar ordering takes place and collective motion sets
in~\cite{vicsek1995novel,toner1995long,gregoire2003moving,Chate:2008is,Weber:2013bj,Bricard:2013jq,Marchetti:2013bp,Peshkov:2014un}.
We are here interested in this aligning behavior at high packing fractions,
where, in the absence of activity, the equilibrium system forms a crystalline
phase.}

\od{Within the present state of knowledge one cannot predict whether the system
spontaneously breaks both the rotational symmetry associated with the
global phase of the polarity and the translational symmetry of space associated
with the structural ordering -- a situation analogous to that of
supersolidity~\cite{Kuklov:2011bl}}. \ms{ On one hand a mesoscopic field
theory~\cite{Menzel:2013gs} predicts a transition from a resting crystal to a
traveling crystalline state, where the particles migrate collectively while
keeping their crystalline order.} \od{On the other hand,
simulations~\cite{Weber:2014cz} of a generalized Vicsek
model~\cite{Gregoire:2004ica}, which accounts for alignment as well as
short-ranged repulsive interactions, report a mutual exclusion of the polar and
the structural order and therefore refutes the generality of the theory.}
\ms{Furthermore, in both cases, the speed of the particles and the alignment
strength -- equivalently the noise level -- are fixed, while, in principle, one
expects the density to renormalize these parameters. The Motility Induced Phase
Separation, for instance, is precisely a consequence of a decrease in particle
speed with increasing local
density~\cite{Farrell:2012vf,Redner:2012vr,Cates:2014dn,
Barre:2014dp,MartinGomez:2018wi}.}
\begin{figure}%
  \centering
  \includegraphics{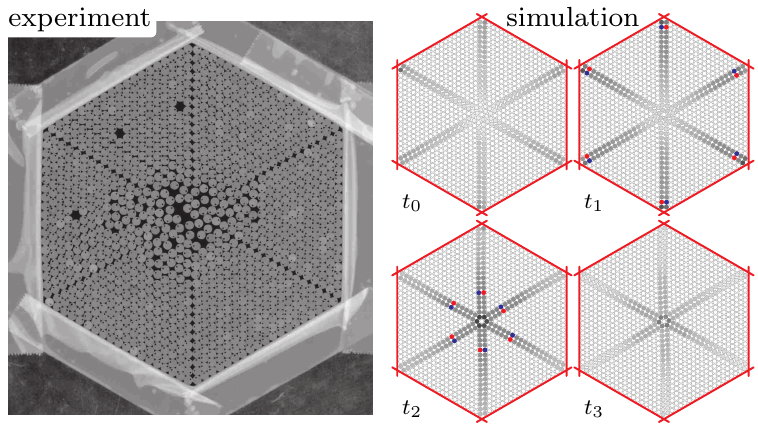}%
  \caption{Flowing crystal configurations, in the experiment at finite noise
  level with geometrical frustration ($N=1104$, $\phi=0.86$), and in
  simulations at successive times, without noise and without geometrical
  frustration; the gray levels code for the orientational order parameter,
  colors blue/red code for 5/7 neighbors; ($N=1141$, $\phi=0.88$).}%
  \label{fig:expnum}%
  \vspace{-5mm}
\end{figure}

\od{For any specific system of interest, it is thus of interest to consider the
following questions. (i)~Does alignment develop at high density, and, if it is
the case, does polar order take place? (ii)~Is the structural order stable
against activity in the presence of polar alignment? For active nematics,
activity promotes the proliferation of defects and destroys nematic
order~\cite{Narayan:2007bg,Sanchez:2012gt,Wensink:2012ci,Giomi:2013ky,Zhou:2014gl}.
(iii)~If the traveling crystalline state exists, how does it adapt to
confinement? In a fluid phase, active flocks condensate at the
boundaries~\cite{Fily:2014gy} and form vortex
flows~\cite{Deseigne:2012kn,Woodhouse:2012vx,Bricard:2015jx,Hiraoka:2017fw}.
How does it translate to a traveling crystalline phase? Does it stop, does it
melt, does it fracture?}
\begin{figure*}%
  \centering
  \includegraphics{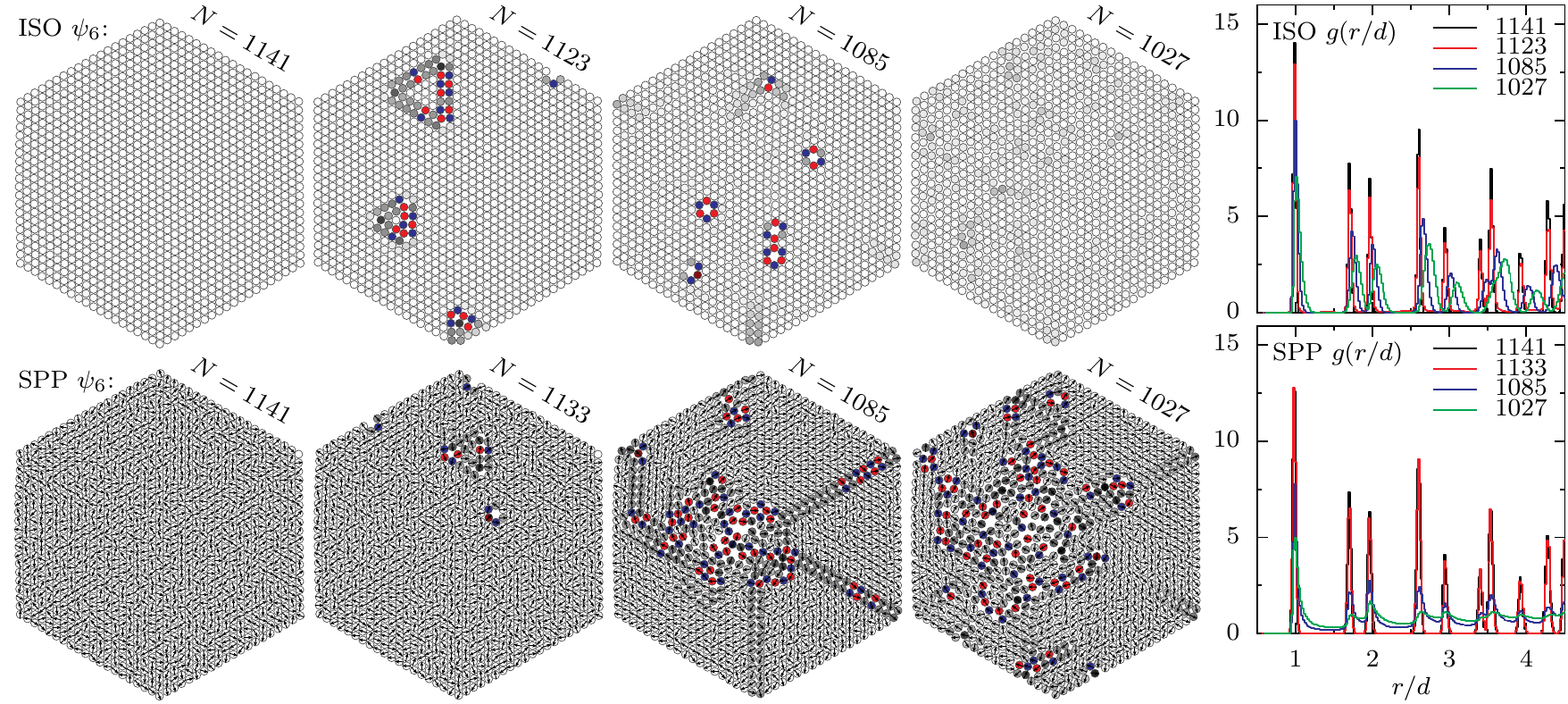}%
  \caption{\textbf{Experimental data}, static aspects: Left panels show
  time-snapshots of isotropic disks (ISO, top row) and of self-propelled polar
  disks (SPP, bottom row). The grayscale indicates individual values of the
  local orientational order parameter $\psi_6\in[0,1]$, except when there are
  more/less than six Voronoi neighbors (red/blue). The rightmost panels show
  the corresponding pair correlation functions, obtained from time-averages
  over trajectories.}%
  \label{fig:static}%
  \vspace{-4mm}%
\end{figure*}%

\od{In this letter we consider a system of self-propelled hard
disks~\cite{Deseigne:2010gc}, for which alignment emerges from the dynamical
relaxation of the particle polarity~\cite{Lam:2015bp} after each collision, and
induces  collective motion at moderate packing
fraction~\cite{Deseigne:2010gc,Weber:2013bj,Lam:2015bp}.}
\od{Building upon this well characterized experimental system, we demonstrate
(i)~that alignment survives at high density, (ii)~that the system develops both
polar and structural ordering on the experimental length-scales and (iii)~that
this polar and ordered structure organizes into a coherent sheared flow, which
is made possible by the localization of shear along intermittent stacking
faults (Fig.~\ref{fig:expnum} and video SI-1). The emergence of shear in
response to boundary conditions is the basic property defining liquids. For
obvious reasons, however, we cannot call this coherently flowing ordered
structure a liquid-crystal; we shall call it a \emph{rheo-crystal}.}
\od{The existence of this rheo-crystal does not rely on the particular
experimental conditions under which it is observed. Simulations very well
reproduce the experimentally observed phenomenology (video SI-2). Within
periodic boundary conditions, they confirm the existence of a traveling ordered
structure (video SI-6). Within the hexagonal geometry, they allow for relaxing
the confinement gradually instead of removing particles and varying the control
parameters. The structurally ordered flow remains coherent for a large set of
parameters and up to the largest scale we could investigate. Finally, the limit
of zero noise  (Fig.~\ref{fig:expnum}-right and video SI-7) allows us to
precise the dynamics along the stacking faults.}

The experimental system is made of micro-machined disks (diameter $d=4$\,mm,
total height $2$\,mm), with an off-center tip and a glued rubber skate, located
at diametrically opposite positions. These two ``legs'' with different
mechanical response endow the particles with a polar axis. Subjected to
vertical vibration, these polar disks (SPP) perform a persistent random walk,
the persistence length of which is set by the vibration parameters (sinusoidal
vibration with frequency $f=95$\,Hz and amplitude~$a$)~\cite{Deseigne:2010gc}.
The maximal acceleration relative to gravity is set to $\Gamma = (2\pi f)^2 a/g
= 2.4$. A glass plate confines the particles from above. We also use plain
rotationally-invariant disks (same metal, diameter, and height), hereafter
called the ``isotropic''~(ISO) disks. In the present study, the disks are
laterally confined in a regular hexagonal arena of side length~$79$\,mm. The
motion of all particles is tracked using a standard CCD~camera at a frame rate
of $30$\,Hz.

\od{As already stated in~\cite{Deseigne:2012kn}, it is possible to fix, but
very difficult to fine tune in a reproducible manner the self propulsion in the
above experimental system. This is why, already in~\cite{Weber:2013bj}, we have
constructed a model for the motion and collisions of the polar disks, which
accounts quantitatively for the experimental properties at the single and pair
interaction level and reproduces very well the macroscopic physics.} In this
model, later studied in great detail~\cite{Lam:2015bp}, a particle~$i$ is
described by position~$\bfr_i$ and velocity~$\bfv_i$ of its center, and by its
body axis given by the unit vector~$\bfn_i$. Between collisions these
parameters evolve according to the equations \vspace{-2mm}
\begin{subequations}
\label{eq:dyn}
\begin{align}
  \label{eq:dotbfr}
  \tfrac{\text{d}}{\text{d}t}\bfr_i &= \bfv_i, \\
  \label{eq:dotbfv}
  \tau_v \tfrac{\text{d}}{\text{d}t}\bfv_i &= \bfn_i - \bfv_i, \\
  \label{eq:dotbfn}
  \tau_n \tfrac{\text{d}}{\text{d}t}\bfn_i &= (\bfn_i \times \hat\bfv_i) \times \bfn_i,
\end{align}
\end{subequations}
which have been made dimensionless by choosing suitable units. In
Eq.~\eqref{eq:dotbfv}, the competition between self-propulsion~$\bfn$ and
viscous damping~$-\bfv$ lets the velocity relax to~$\bfn$ on a
timescale~$\tau_v$.
\begin{figure*}[t!]%
  \centering
  \includegraphics{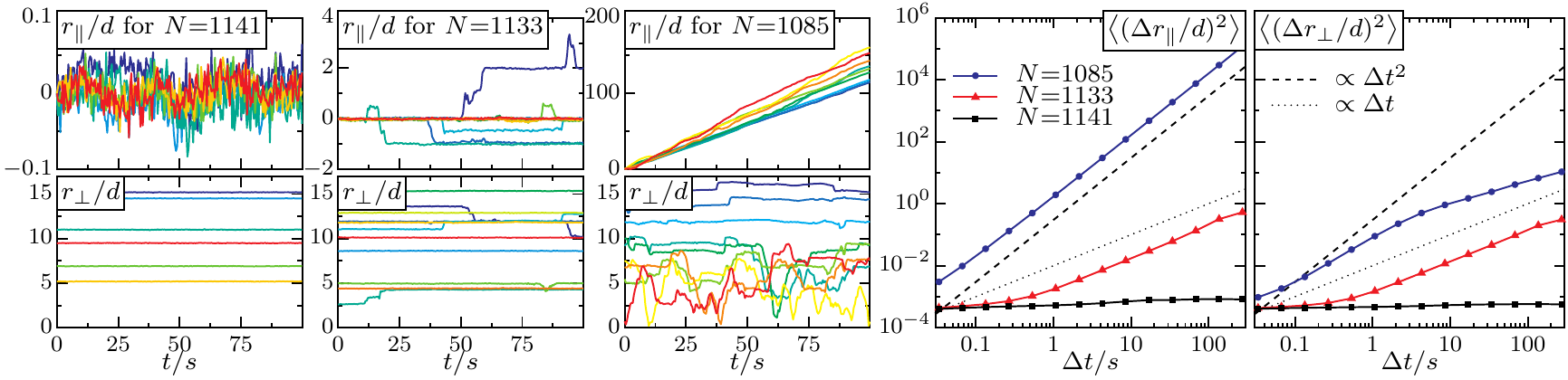}%
  \vspace{-3mm}%
  \caption{\textbf{Experimental data}, SPP dynamics: The panels on the left
  show typical trajectories for different~$N$. Plotted are the perpendicular
  and (unwrapped) parallel components of the particle position~\cite{Note1} The
  panels on the right show the corresponding mean squared displacements.}%
  \label{fig:spponly}%
  \vspace{-4mm}%
\end{figure*}
Similarly, in Eq.~\eqref{eq:dotbfn}, the polarity $\bfn$ undergoes an
overdamped torque that orients it toward~$\bfv$ on a timescale~$\tau_n$.
Interactions between particles are elastic hard-disk collisions which change
$\bfv$ but not~$\bfn$.  After such a collision, $\bfv$~and $\bfn$ are not
collinear, and the particles undergo curved trajectories which either are
interrupted by another collision, or the particles reach their stationary
state, characterized by a straight trajectory at unit speed $\bfv=\bfn$. The
final direction of $\bfv$ (and of~$\bfn$) depends on the parameter
$\alpha=\tau_n/\tau_v$, which can be understood as the persistence of the
polarity~$\bfn$. On top of the deterministic trajectories given by
Eqs.~\eqref{eq:dyn}, we add to both vectors a common angular noise, normally
distributed with zero mean and variance~$2D$ (see~\cite{Lam:2015bp} for
implementation details). In the following, \od{unless specified}, only the size
of the arena and the number of particles are varied; the parameters
$\tau_v=0.25$, $\alpha=0.867$, $D=2.16$ have been chosen as to be in the
experimental range. The hexagonal packing is qualified using the local
orientational order parameter $\psi_6 :=
\bigl|\sum_q\exp\bigl(6i\theta_{pq}\bigr)\bigr|\bigm/n_p$, defined as a
per-particle quantity. Here, the sum runs over the $n_p$~neighbors of
particle~$p$, and $\theta_{pq}$ is the angle of the connecting vector, with
respect to a fixed axis. Particle neighborship is defined by Voronoi
tessellation.

\od{We start with the description of the experimental results, at the root of
the present work.} Figure~\ref{fig:static} reports some static aspects of the
structures obtained in the experiment. The densest hexagonal packing we can
form in the arena ($N=1141$, $\phi=0.89$) is first prepared, before we remove
particles at random. The defectless structures (leftmost column) are stable,
both for the ISO and SPP disks. In the case of the ISO disks (top row),
removing a few particles creates defects, which locally lower the individual
order parameter~$\psi_6$ (panels $N=1123,1085$). When the number of removed
particles corresponds exactly to the outermost layer of the hexagonal packing
(here $114$ particles, passing to $N=1027$), one recovers a crystal of ISO
disks without geometric frustration and still far from melting ($\phi=0.801$).

By contrast, in the case of SPP disks (bottom row) the structure becomes
increasingly disordered, and defects proliferate. This disorder is reflected in
the shape of the pair correlation functions (rightmost column in
Fig.~\ref{fig:static}) which do not drop to zero between peaks. A remarkable
feature is the existence of highly ordered sectors, which are separated by less
dense pairs of lines where the local symmetry is square ($N=1085$). One further
notices an increasingly large zone \ms{of lower density} in the center of the
arena ($N=1085,1027$).

The three regimes foreshadowed in the bottom row of Fig.~\ref{fig:static}
($N=1141,1133,1085$) also differ in their dynamics, as can be seen from
Fig.~\ref{fig:spponly} and from videos~SI-3 to SI-5. For the perfect hexagonal
packing, the SPP disks remain trapped at the same average position, and the
corresponding mean square displacements~(MSD) are that of a frozen structure
(see video SI-3). For $N=1133$ one observes jumps in the particle trajectories
both in the parallel and perpendicular directions~\footnote{We used six
different Cartesian coordinate systems in the six sectors of the hexagon,
aligned with the respective boundaries. This allows to speak of parallel and
perpendicular coordinates, $r_\parallel, r_\perp$.}: the particles exchange
neighbors, change layers and overall diffuse in both directions as evidenced by
the linear increase with time of their~MSD. Video~SI-4 reveals the existence of
small but persistent depleted regions, or loose ``droplets'', which rapidly
explore all the system. These droplets are composed of vacancies and defects,
which condensate the volume left by the removed particles. Such droplets
already are complex objects, which split and merge, and the motion of which
does not reduce to that of a specific defect. In this regime, we are thus in
presence of a crystalline structure, the density fluctuations of which relax on
fast timescales. This finding extends the previously reported decoupling
between structure and dynamics~\cite{Briand:2016fj} to very large packing
fraction. Finally, for $N=1085$, one observes ballistic-like displacement along
the direction parallel to the walls, while the perpendicular motion remains
diffusive. Video~SI-5 shows how the whole crystalline structure starts flowing
and forms a global rotation, responsible for the reported ballistic motion.
This flowing crystalline structure is made possible by the condensation of
shear along localized stacking faults, dividing the hexagon in regular
triangles. Along these lines the local symmetry oscillates between hexagonal
and square, as the blocks slide past each other.

\od{The above results clearly answer the questions raised in the introduction.
In particular they demonstrate that the dynamical alignment survives the
high-frequency collisions and triggers collective motions, which take the form
of a structurally ordered flow. Accordingly, one expects that, within periodic
boundaries, a traveling crystal will form, as we confirm with numerical
simulations (see video SI-6). They also raise a number of new questions.
(i)~Removing particles, we induce geometrical frustration. Is it necessary for
the global rotation to develop? (ii)~While the boundaries facilitate the
crystalline order, defects concentrate in the hexagon center, where the
rotating flow does not develop. What if the system size is increased?}
\ms{(iii)~How do parameters such as density or noise level impact on the
rheo-crystal?} To answer these questions, we now turn to numerical simulations
of the model introduced above.

\begin{figure}[t!]%
  \centering
  \includegraphics{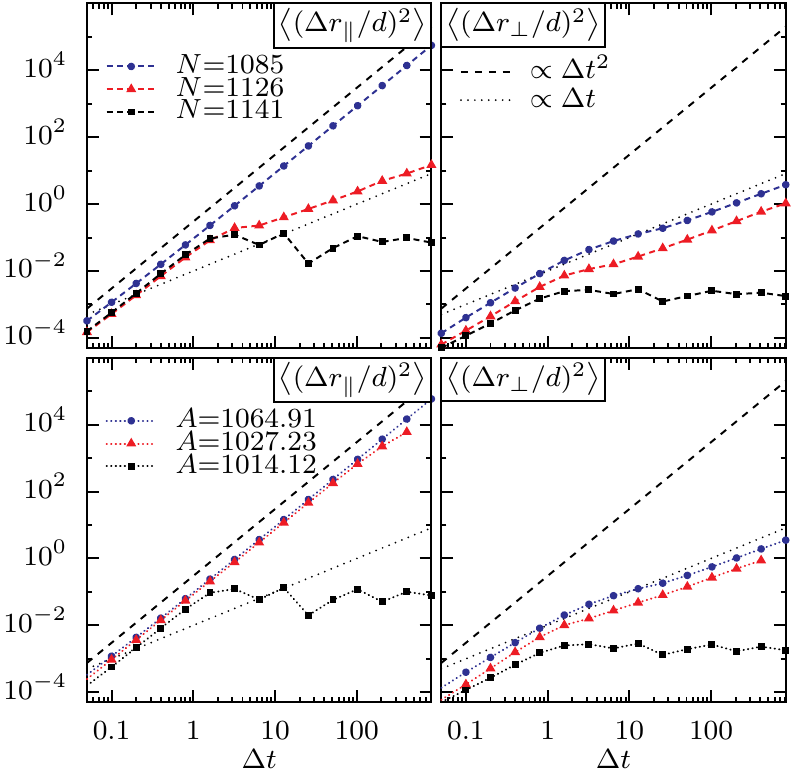}%
  \vspace{-3mm}%
  \caption{\textbf{Simulation data}: Variation of~$N$ (top row) versus
  variation of the arena area~$A$ (bottom row). Shown are mean squared
  displacements obtained from pairs $(N,A)$ with the common (bulk) volume
  fractions $\phi=0.846$, $0.878$, and $0.890$.}%
  \label{fig:msd_sim}%
\end{figure}

\od{First, the global motion is not induced by geometric frustration.
Instead of removing particles, we target the same packing fraction without
geometric frustration by increasing the arena area~$A$.
Figure~\ref{fig:msd_sim} displays the parallel and perpendicular~MSD for the
SPP disks, once obtained by removing particles (top row) and once by
increasing~$A$ (bottom row). The packing fractions are the same in both rows,
but the dynamical behavior is not: the parallel diffusion at $N=1126$ is
replaced by rotation at the corresponding $A=1027.23$. Removing frustration
actually favors the coherent flow of the crystalline structure.} \ms{It is the
droplet regime, where both MSD are diffusive, which is absent in the absence of
frustration. We further learn that the central region of lowered density and
order, which is still present in the absence of frustration, does not result
from the coalescence of such droplets. It is rather a consequence of the close
packing of the particles along the hexagon boundaries.}
\begin{figure}%
  \centering
  \includegraphics{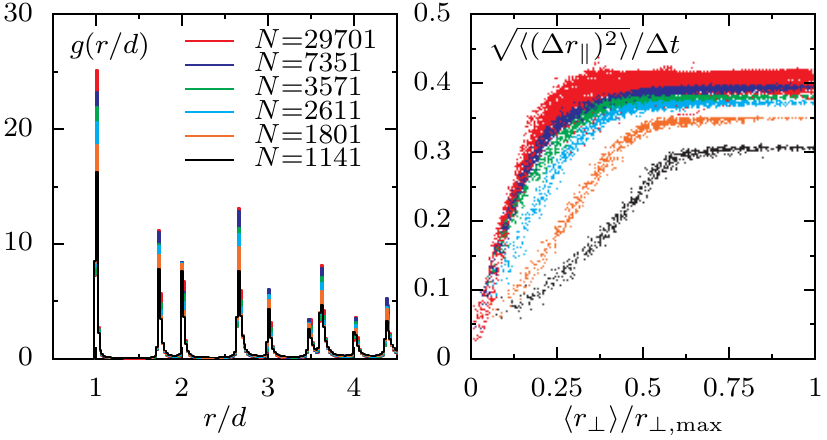}%
  \vspace{-4mm}%
  \caption{{\bf Simulation data} from systems of different sizes. Left:
  pair-correlation functions, right: average displacement per unit time
  vs.~distance from the center rescaled by the system size. ($A=1027.23$,
  $\phi=0.878$, $\Delta t=256$)}%
  \label{fig:gr_sim}%
  \vspace{-4mm}%
\end{figure}

\ms{Second, as summarized in Fig.~\ref{fig:gr_sim}, the larger the system is,
the better defined is its structure and the faster it flows: Larger systems
have higher principal peaks in the pair correlation function (left panel).
Regarding the dynamics, the right panel of Fig.~\ref{fig:gr_sim} displays the
average parallel displacement per unit time as a function of the rescaled
average distance from the center. With increasing size, the rotation becomes
more important in two ways, the overall magnitude of the rotation increases,
and the central region, where the rotation is reduced, occupies a smaller
fraction of the system. In all simulations used for Fig.~\ref{fig:gr_sim}, the
vortex flow organizes at the scale of the system size. From that point of view,
alignment actually promotes system-size correlations in the
structure~\footnote{Evidently, the presence of the boundaries dominates the
global properties of flow and structure. This may be different for much larger
systems, but this is pure speculation. We focus on finite systems, where the
importance of boundaries even helps to keep the corresponding equilibrium
system crystalline.}. The breaking of structure, necessary to allow global
flow, happens only along the fault lines, where particles temporarily adopt a
local square symmetry. These lines become less important with increasing system
size, since they scale as length, not as area.}
\begin{figure}[t!]%
  \centering
  \includegraphics{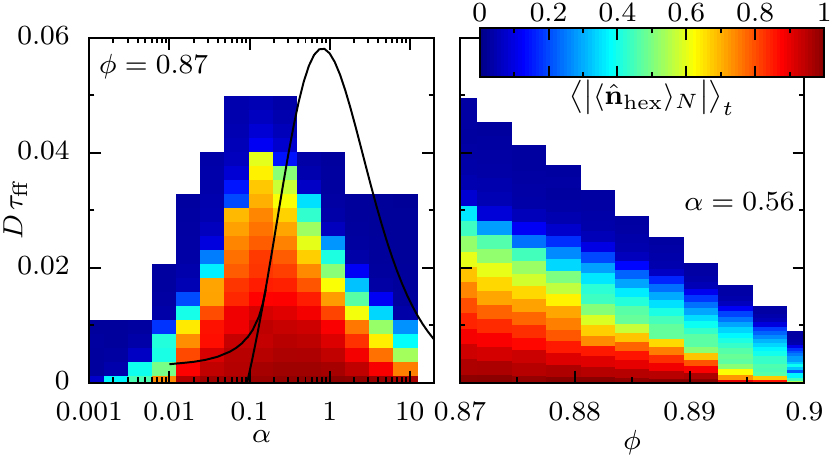}%
  \vspace{-3mm}%
  \caption{\textbf{Simulation data}, exploring the parameter space in the
  $(\alpha,D)$ and $(\phi,D)$ planes. The color codes for coincidence of
  individual orientation vectors with the orientations of global flow, first
  averaged over the particles, then over time. Noise levels are scaled by the
  free-flight time~$\tau_{\text{ff}}(\phi)$, as measured at equilibrium. The
  curves in the left panel are the disorder--polar transition lines in the
  dilute limit, taken from Ref.~\cite{Lam:2015bp}, but scaled by~$0.2$ for
  reasons of visual presentation.}%
  \label{fig:phasediag}%
  \vspace{-4mm}
\end{figure}

\ms{Third, as shown in Fig.~\ref{fig:phasediag}, the phase diagram is
strikingly similar to the one observed in the limit of vanishing densities (see
the curves in Fig.~\ref{fig:phasediag}-left, also Fig.~2c in
Ref.~\cite{Lam:2015bp}). For strong noise no collective motion takes place and
a static, but active, crystal is stabilized. For weak noise, the rheo-crystal
occupies a significant range of~$\alpha$. As in the dilute case (curves in
Fig.~\ref{fig:phasediag}-left) the flowing region has a peak-like shape; for
moderate noise, too small or too large polarity persistence hinders collective
alignment. Note the non-trivial influence of the density: on the large-$\alpha$
side of the peak, density promotes disorder (Fig.~\ref{fig:phasediag}-right),
while it contributes to alignment on the low-$\alpha$ side.}

\od{Finally, taking advantage of the fact that the rheo-crystal is present in
the absence of noise, we precise the dynamics taking place along the stacking
fault lines (see video~SI-7).} \ms{The lattices in the different sectors of the
hexagon match seamlessly only at specific times (Fig~\ref{fig:expnum} at
$t_3$). At most times, the active flow distorts the structure, very locally,
along the stacking fault lines. This starts with the emission of dislocations
from the boundary corners. They rapidly travel towards the center, where they
merge and vanish. Note however that the mismatch across these fault lines is a
displacement of the hexagonal lattice, not a rotation. We therefore do not
think of them as grain boundaries.}

\od{Altogether, the present experimental work, supported by numerical
simulations, has unveiled a new \ms{behavior} of active matter, which had not
been observed before. Let us conclude by stressing that many open question
remain, especially regarding the precise nature of the traveling crystal phase,
in the large $N$ limit. Further intensive numerical simulations will be
necessary to address these issues.}

{\it --- Acknowledgment ---} We acknowledge financial support from \emph{Ecole
Doctorale~ED564 ``Physique en Ile~de France''} for Guillaume Briand's PhD
grant. We thank Vincenzo Vitelli and Pierre Ronceray for interesting
discussions.

\bibliography{rheocrystal}

\begin{thebibliography}{34}%
\makeatletter
\providecommand \@ifxundefined [1]{%
 \@ifx{#1\undefined}
}%
\providecommand \@ifnum [1]{%
 \ifnum #1\expandafter \@firstoftwo
 \else \expandafter \@secondoftwo
 \fi
}%
\providecommand \@ifx [1]{%
 \ifx #1\expandafter \@firstoftwo
 \else \expandafter \@secondoftwo
 \fi
}%
\providecommand \natexlab [1]{#1}%
\providecommand \enquote  [1]{``#1''}%
\providecommand \bibnamefont  [1]{#1}%
\providecommand \bibfnamefont [1]{#1}%
\providecommand \citenamefont [1]{#1}%
\providecommand \href@noop [0]{\@secondoftwo}%
\providecommand \href [0]{\begingroup \@sanitize@url \@href}%
\providecommand \@href[1]{\@@startlink{#1}\@@href}%
\providecommand \@@href[1]{\endgroup#1\@@endlink}%
\providecommand \@sanitize@url [0]{\catcode `\\12\catcode `\$12\catcode
  `\&12\catcode `\#12\catcode `\^12\catcode `\_12\catcode `\%12\relax}%
\providecommand \@@startlink[1]{}%
\providecommand \@@endlink[0]{}%
\providecommand \url  [0]{\begingroup\@sanitize@url \@url }%
\providecommand \@url [1]{\endgroup\@href {#1}{\urlprefix }}%
\providecommand \urlprefix  [0]{URL }%
\providecommand \Eprint [0]{\href }%
\providecommand \doibase [0]{http://dx.doi.org/}%
\providecommand \selectlanguage [0]{\@gobble}%
\providecommand \bibinfo  [0]{\@secondoftwo}%
\providecommand \bibfield  [0]{\@secondoftwo}%
\providecommand \translation [1]{[#1]}%
\providecommand \BibitemOpen [0]{}%
\providecommand \bibitemStop [0]{}%
\providecommand \bibitemNoStop [0]{.\EOS\space}%
\providecommand \EOS [0]{\spacefactor3000\relax}%
\providecommand \BibitemShut  [1]{\csname bibitem#1\endcsname}%
\let\auto@bib@innerbib\@empty
\bibitem [{\citenamefont {Bialk{\'e}}\ \emph {et~al.}(2012)\citenamefont
  {Bialk{\'e}}, \citenamefont {Speck},\ and\ \citenamefont
  {L{\"o}wen}}]{Bialke:2012cw}%
  \BibitemOpen
  \bibfield  {author} {\bibinfo {author} {\bibfnamefont {J.}~\bibnamefont
  {Bialk{\'e}}}, \bibinfo {author} {\bibfnamefont {T.}~\bibnamefont {Speck}}, \
  and\ \bibinfo {author} {\bibfnamefont {H.}~\bibnamefont {L{\"o}wen}},\
  }\href@noop {} {\bibfield  {journal} {\bibinfo  {journal} {PRL}\ }\textbf
  {\bibinfo {volume} {108}},\ \bibinfo {pages} {168301} (\bibinfo {year}
  {2012})}\BibitemShut {NoStop}%
\bibitem [{\citenamefont {Klamser}\ \emph {et~al.}(2018)\citenamefont
  {Klamser}, \citenamefont {Kapfer},\ and\ \citenamefont
  {Krauth}}]{Klamser:2018vk}%
  \BibitemOpen
  \bibfield  {author} {\bibinfo {author} {\bibfnamefont {J.~U.}\ \bibnamefont
  {Klamser}}, \bibinfo {author} {\bibfnamefont {S.~C.}\ \bibnamefont {Kapfer}},
  \ and\ \bibinfo {author} {\bibfnamefont {W.}~\bibnamefont {Krauth}},\
  }\href@noop {} {\bibfield  {journal} {\bibinfo  {journal} {arXiv}\ }
  (\bibinfo {year} {2018})},\ \Eprint {http://arxiv.org/abs/1802.10021v1}
  {1802.10021v1} \BibitemShut {NoStop}%
\bibitem [{\citenamefont {Vicsek}\ \emph {et~al.}(1995)\citenamefont {Vicsek},
  \citenamefont {Czir{\'o}k}, \citenamefont {Ben-Jacob}, \citenamefont
  {Cohen},\ and\ \citenamefont {Shochet}}]{vicsek1995novel}%
  \BibitemOpen
  \bibfield  {author} {\bibinfo {author} {\bibfnamefont {T.}~\bibnamefont
  {Vicsek}}, \bibinfo {author} {\bibfnamefont {A.}~\bibnamefont {Czir{\'o}k}},
  \bibinfo {author} {\bibfnamefont {E.}~\bibnamefont {Ben-Jacob}}, \bibinfo
  {author} {\bibfnamefont {I.}~\bibnamefont {Cohen}}, \ and\ \bibinfo {author}
  {\bibfnamefont {O.}~\bibnamefont {Shochet}},\ }\href@noop {} {\bibfield
  {journal} {\bibinfo  {journal} {Phys. Rev. Lett.}\ }\textbf {\bibinfo
  {volume} {75}},\ \bibinfo {pages} {1226} (\bibinfo {year}
  {1995})}\BibitemShut {NoStop}%
\bibitem [{\citenamefont {Toner}\ and\ \citenamefont
  {Tu}(1995)}]{toner1995long}%
  \BibitemOpen
  \bibfield  {author} {\bibinfo {author} {\bibfnamefont {J.}~\bibnamefont
  {Toner}}\ and\ \bibinfo {author} {\bibfnamefont {Y.}~\bibnamefont {Tu}},\
  }\href@noop {} {\bibfield  {journal} {\bibinfo  {journal} {Phys. Rev. Lett.}\
  }\textbf {\bibinfo {volume} {75}},\ \bibinfo {pages} {4326} (\bibinfo {year}
  {1995})}\BibitemShut {NoStop}%
\bibitem [{\citenamefont {Gr{\'e}goire}\ \emph {et~al.}(2003)\citenamefont
  {Gr{\'e}goire}, \citenamefont {Chat{\'e}},\ and\ \citenamefont
  {Tu}}]{gregoire2003moving}%
  \BibitemOpen
  \bibfield  {author} {\bibinfo {author} {\bibfnamefont {G.}~\bibnamefont
  {Gr{\'e}goire}}, \bibinfo {author} {\bibfnamefont {H.}~\bibnamefont
  {Chat{\'e}}}, \ and\ \bibinfo {author} {\bibfnamefont {Y.}~\bibnamefont
  {Tu}},\ }\href@noop {} {\bibfield  {journal} {\bibinfo  {journal} {Physica D:
  Nonlinear Phenomena}\ }\textbf {\bibinfo {volume} {181}},\ \bibinfo {pages}
  {157} (\bibinfo {year} {2003})}\BibitemShut {NoStop}%
\bibitem [{\citenamefont {Chat{\'e}}\ \emph {et~al.}(2008)\citenamefont
  {Chat{\'e}}, \citenamefont {Ginelli}, \citenamefont {Gr{\'e}goire},\ and\
  \citenamefont {Raynaud}}]{Chate:2008is}%
  \BibitemOpen
  \bibfield  {author} {\bibinfo {author} {\bibfnamefont {H.}~\bibnamefont
  {Chat{\'e}}}, \bibinfo {author} {\bibfnamefont {F.}~\bibnamefont {Ginelli}},
  \bibinfo {author} {\bibfnamefont {G.}~\bibnamefont {Gr{\'e}goire}}, \ and\
  \bibinfo {author} {\bibfnamefont {F.}~\bibnamefont {Raynaud}},\ }\href@noop
  {} {\bibfield  {journal} {\bibinfo  {journal} {Phys. Rev. E}\ }\textbf
  {\bibinfo {volume} {77}},\ \bibinfo {pages} {046113} (\bibinfo {year}
  {2008})}\BibitemShut {NoStop}%
\bibitem [{\citenamefont {Weber}\ \emph {et~al.}(2013)\citenamefont {Weber},
  \citenamefont {Hanke}, \citenamefont {Deseigne}, \citenamefont {L{\'e}onard},
  \citenamefont {Dauchot}, \citenamefont {Frey},\ and\ \citenamefont
  {Chat{\'e}}}]{Weber:2013bj}%
  \BibitemOpen
  \bibfield  {author} {\bibinfo {author} {\bibfnamefont {C.~A.}\ \bibnamefont
  {Weber}}, \bibinfo {author} {\bibfnamefont {T.}~\bibnamefont {Hanke}},
  \bibinfo {author} {\bibfnamefont {J.}~\bibnamefont {Deseigne}}, \bibinfo
  {author} {\bibfnamefont {S.}~\bibnamefont {L{\'e}onard}}, \bibinfo {author}
  {\bibfnamefont {O.}~\bibnamefont {Dauchot}}, \bibinfo {author} {\bibfnamefont
  {E.}~\bibnamefont {Frey}}, \ and\ \bibinfo {author} {\bibfnamefont
  {H.}~\bibnamefont {Chat{\'e}}},\ }\href@noop {} {\bibfield  {journal}
  {\bibinfo  {journal} {Phys. Rev. Lett.}\ }\textbf {\bibinfo {volume} {110}},\
  \bibinfo {pages} {208001} (\bibinfo {year} {2013})}\BibitemShut {NoStop}%
\bibitem [{\citenamefont {Bricard}\ \emph {et~al.}(2013)\citenamefont
  {Bricard}, \citenamefont {Caussin}, \citenamefont {Desreumaux}, \citenamefont
  {Dauchot},\ and\ \citenamefont {Bartolo}}]{Bricard:2013jq}%
  \BibitemOpen
  \bibfield  {author} {\bibinfo {author} {\bibfnamefont {A.}~\bibnamefont
  {Bricard}}, \bibinfo {author} {\bibfnamefont {J.-B.}\ \bibnamefont
  {Caussin}}, \bibinfo {author} {\bibfnamefont {N.}~\bibnamefont {Desreumaux}},
  \bibinfo {author} {\bibfnamefont {O.}~\bibnamefont {Dauchot}}, \ and\
  \bibinfo {author} {\bibfnamefont {D.}~\bibnamefont {Bartolo}},\ }\href@noop
  {} {\bibfield  {journal} {\bibinfo  {journal} {Nature}\ }\textbf {\bibinfo
  {volume} {503}},\ \bibinfo {pages} {95} (\bibinfo {year} {2013})}\BibitemShut
  {NoStop}%
\bibitem [{\citenamefont {Marchetti}\ \emph {et~al.}(2013)\citenamefont
  {Marchetti}, \citenamefont {Joanny}, \citenamefont {Ramaswamy}, \citenamefont
  {Liverpool}, \citenamefont {Prost}, \citenamefont {Rao},\ and\ \citenamefont
  {Simha}}]{Marchetti:2013bp}%
  \BibitemOpen
  \bibfield  {author} {\bibinfo {author} {\bibfnamefont {M.~C.}\ \bibnamefont
  {Marchetti}}, \bibinfo {author} {\bibfnamefont {J.-F.}\ \bibnamefont
  {Joanny}}, \bibinfo {author} {\bibfnamefont {S.}~\bibnamefont {Ramaswamy}},
  \bibinfo {author} {\bibfnamefont {T.~B.}\ \bibnamefont {Liverpool}}, \bibinfo
  {author} {\bibfnamefont {J.}~\bibnamefont {Prost}}, \bibinfo {author}
  {\bibfnamefont {M.}~\bibnamefont {Rao}}, \ and\ \bibinfo {author}
  {\bibfnamefont {R.~A.}\ \bibnamefont {Simha}},\ }\href@noop {} {\bibfield
  {journal} {\bibinfo  {journal} {Rev. Mod. Phys.}\ }\textbf {\bibinfo {volume}
  {85}},\ \bibinfo {pages} {1143} (\bibinfo {year} {2013})}\BibitemShut
  {NoStop}%
\bibitem [{\citenamefont {Peshkov}\ \emph {et~al.}(2014)\citenamefont
  {Peshkov}, \citenamefont {Bertin}, \citenamefont {Ginelli},\ and\
  \citenamefont {Chat{\'e}}}]{Peshkov:2014un}%
  \BibitemOpen
  \bibfield  {author} {\bibinfo {author} {\bibfnamefont {A.}~\bibnamefont
  {Peshkov}}, \bibinfo {author} {\bibfnamefont {E.~M.}\ \bibnamefont {Bertin}},
  \bibinfo {author} {\bibfnamefont {F.}~\bibnamefont {Ginelli}}, \ and\
  \bibinfo {author} {\bibfnamefont {H.}~\bibnamefont {Chat{\'e}}},\ }\href@noop
  {} {\bibfield  {journal} {\bibinfo  {journal} {The European Physical
  Journal-Special Topics}\ }\textbf {\bibinfo {volume} {223}},\ \bibinfo
  {pages} {1315} (\bibinfo {year} {2014})}\BibitemShut {NoStop}%
\bibitem [{\citenamefont {Kuklov}\ \emph {et~al.}(2011)\citenamefont {Kuklov},
  \citenamefont {Prokof{\textquoteright}ev},\ and\ \citenamefont
  {Svistunov}}]{Kuklov:2011bl}%
  \BibitemOpen
  \bibfield  {author} {\bibinfo {author} {\bibfnamefont {A.}~\bibnamefont
  {Kuklov}}, \bibinfo {author} {\bibfnamefont {N.}~\bibnamefont
  {Prokof{\textquoteright}ev}}, \ and\ \bibinfo {author} {\bibfnamefont
  {B.}~\bibnamefont {Svistunov}},\ }\href@noop {} {\bibfield  {journal}
  {\bibinfo  {journal} {Physics}\ }\textbf {\bibinfo {volume} {4}},\ \bibinfo
  {pages} {1107} (\bibinfo {year} {2011})}\BibitemShut {NoStop}%
\bibitem [{\citenamefont {Menzel}\ and\ \citenamefont
  {L{\"o}wen}(2013)}]{Menzel:2013gs}%
  \BibitemOpen
  \bibfield  {author} {\bibinfo {author} {\bibfnamefont {A.~M.}\ \bibnamefont
  {Menzel}}\ and\ \bibinfo {author} {\bibfnamefont {H.}~\bibnamefont
  {L{\"o}wen}},\ }\href@noop {} {\bibfield  {journal} {\bibinfo  {journal}
  {PRL}\ }\textbf {\bibinfo {volume} {110}},\ \bibinfo {pages} {055702}
  (\bibinfo {year} {2013})}\BibitemShut {NoStop}%
\bibitem [{\citenamefont {Weber}\ \emph {et~al.}(2014)\citenamefont {Weber},
  \citenamefont {Bock},\ and\ \citenamefont {Frey}}]{Weber:2014cz}%
  \BibitemOpen
  \bibfield  {author} {\bibinfo {author} {\bibfnamefont {C.~A.}\ \bibnamefont
  {Weber}}, \bibinfo {author} {\bibfnamefont {C.}~\bibnamefont {Bock}}, \ and\
  \bibinfo {author} {\bibfnamefont {E.}~\bibnamefont {Frey}},\ }\href@noop {}
  {\bibfield  {journal} {\bibinfo  {journal} {PRL}\ }\textbf {\bibinfo {volume}
  {112}},\ \bibinfo {pages} {168301} (\bibinfo {year} {2014})}\BibitemShut
  {NoStop}%
\bibitem [{\citenamefont {Gr{\'e}goire}\ and\ \citenamefont
  {Chat{\'e}}(2004)}]{Gregoire:2004ica}%
  \BibitemOpen
  \bibfield  {author} {\bibinfo {author} {\bibfnamefont {G.}~\bibnamefont
  {Gr{\'e}goire}}\ and\ \bibinfo {author} {\bibfnamefont {H.}~\bibnamefont
  {Chat{\'e}}},\ }\href@noop {} {\bibfield  {journal} {\bibinfo  {journal}
  {Phys. Rev. Lett.}\ }\textbf {\bibinfo {volume} {92}},\ \bibinfo {pages}
  {025702} (\bibinfo {year} {2004})}\BibitemShut {NoStop}%
\bibitem [{\citenamefont {Farrell}\ \emph {et~al.}(2012)\citenamefont
  {Farrell}, \citenamefont {Marchetti}, \citenamefont {Marenduzzo},\ and\
  \citenamefont {Tailleur}}]{Farrell:2012vf}%
  \BibitemOpen
  \bibfield  {author} {\bibinfo {author} {\bibfnamefont {F.~D.~C.}\
  \bibnamefont {Farrell}}, \bibinfo {author} {\bibfnamefont {M.~C.}\
  \bibnamefont {Marchetti}}, \bibinfo {author} {\bibfnamefont {D.}~\bibnamefont
  {Marenduzzo}}, \ and\ \bibinfo {author} {\bibfnamefont {J.}~\bibnamefont
  {Tailleur}},\ }\href@noop {} {\bibfield  {journal} {\bibinfo  {journal}
  {Phys. Rev. Lett.}\ }\textbf {\bibinfo {volume} {108}},\ \bibinfo {pages}
  {248101} (\bibinfo {year} {2012})}\BibitemShut {NoStop}%
\bibitem [{\citenamefont {Redner}\ \emph {et~al.}(2013)\citenamefont {Redner},
  \citenamefont {Hagan},\ and\ \citenamefont {Baskaran}}]{Redner:2012vr}%
  \BibitemOpen
  \bibfield  {author} {\bibinfo {author} {\bibfnamefont {G.~S.}\ \bibnamefont
  {Redner}}, \bibinfo {author} {\bibfnamefont {M.~F.}\ \bibnamefont {Hagan}}, \
  and\ \bibinfo {author} {\bibfnamefont {A.}~\bibnamefont {Baskaran}},\
  }\href@noop {} {\bibfield  {journal} {\bibinfo  {journal} {Phys. Rev. Lett.}\
  }\textbf {\bibinfo {volume} {110}},\ \bibinfo {pages} {055701} (\bibinfo
  {year} {2013})}\BibitemShut {NoStop}%
\bibitem [{\citenamefont {Cates}\ and\ \citenamefont
  {Tailleur}(2015)}]{Cates:2014dn}%
  \BibitemOpen
  \bibfield  {author} {\bibinfo {author} {\bibfnamefont {M.~E.}\ \bibnamefont
  {Cates}}\ and\ \bibinfo {author} {\bibfnamefont {J.}~\bibnamefont
  {Tailleur}},\ }\href@noop {} {\bibfield  {journal} {\bibinfo  {journal}
  {Annual Review of Condensed Matter Physics, Vol 6}\ }\textbf {\bibinfo
  {volume} {6}},\ \bibinfo {pages} {219} (\bibinfo {year} {2015})}\BibitemShut
  {NoStop}%
\bibitem [{\citenamefont {Barre}\ \emph {et~al.}(2014)\citenamefont {Barre},
  \citenamefont {Chetrite}, \citenamefont {Muratori},\ and\ \citenamefont
  {Peruani}}]{Barre:2014dp}%
  \BibitemOpen
  \bibfield  {author} {\bibinfo {author} {\bibfnamefont {J.}~\bibnamefont
  {Barre}}, \bibinfo {author} {\bibfnamefont {R.}~\bibnamefont {Chetrite}},
  \bibinfo {author} {\bibfnamefont {M.}~\bibnamefont {Muratori}}, \ and\
  \bibinfo {author} {\bibfnamefont {F.}~\bibnamefont {Peruani}},\ }\href@noop
  {} {\bibfield  {journal} {\bibinfo  {journal} {J Stat Phys}\ }\textbf
  {\bibinfo {volume} {158}},\ \bibinfo {pages} {589} (\bibinfo {year}
  {2014})}\BibitemShut {NoStop}%
\bibitem [{\citenamefont {Mart{\'\i}n-G{\'o}mez}\ \emph
  {et~al.}(2018)\citenamefont {Mart{\'\i}n-G{\'o}mez}, \citenamefont {Levis},
  \citenamefont {D{\'\i}az-Guilera},\ and\ \citenamefont
  {Pagonabarraga}}]{MartinGomez:2018wi}%
  \BibitemOpen
  \bibfield  {author} {\bibinfo {author} {\bibfnamefont {A.}~\bibnamefont
  {Mart{\'\i}n-G{\'o}mez}}, \bibinfo {author} {\bibfnamefont {D.}~\bibnamefont
  {Levis}}, \bibinfo {author} {\bibfnamefont {A.}~\bibnamefont
  {D{\'\i}az-Guilera}}, \ and\ \bibinfo {author} {\bibfnamefont
  {I.}~\bibnamefont {Pagonabarraga}},\ }\href@noop {} {\bibfield  {journal}
  {\bibinfo  {journal} {arXiv}\ } (\bibinfo {year} {2018})},\ \Eprint
  {http://arxiv.org/abs/1801.01002v2} {1801.01002v2} \BibitemShut {NoStop}%
\bibitem [{\citenamefont {Narayan}\ \emph {et~al.}(2007)\citenamefont
  {Narayan}, \citenamefont {Ramaswamy},\ and\ \citenamefont
  {Menon}}]{Narayan:2007bg}%
  \BibitemOpen
  \bibfield  {author} {\bibinfo {author} {\bibfnamefont {V.}~\bibnamefont
  {Narayan}}, \bibinfo {author} {\bibfnamefont {S.}~\bibnamefont {Ramaswamy}},
  \ and\ \bibinfo {author} {\bibfnamefont {N.}~\bibnamefont {Menon}},\
  }\href@noop {} {\bibfield  {journal} {\bibinfo  {journal} {Science}\ }\textbf
  {\bibinfo {volume} {317}},\ \bibinfo {pages} {105} (\bibinfo {year}
  {2007})}\BibitemShut {NoStop}%
\bibitem [{\citenamefont {Sanchez}\ \emph {et~al.}(2012)\citenamefont
  {Sanchez}, \citenamefont {Chen}, \citenamefont {DeCamp}, \citenamefont
  {Heymann},\ and\ \citenamefont {Dogic}}]{Sanchez:2012gt}%
  \BibitemOpen
  \bibfield  {author} {\bibinfo {author} {\bibfnamefont {T.}~\bibnamefont
  {Sanchez}}, \bibinfo {author} {\bibfnamefont {D.~T.~N.}\ \bibnamefont
  {Chen}}, \bibinfo {author} {\bibfnamefont {S.~J.}\ \bibnamefont {DeCamp}},
  \bibinfo {author} {\bibfnamefont {M.}~\bibnamefont {Heymann}}, \ and\
  \bibinfo {author} {\bibfnamefont {Z.}~\bibnamefont {Dogic}},\ }\href@noop {}
  {\bibfield  {journal} {\bibinfo  {journal} {Nature}\ }\textbf {\bibinfo
  {volume} {491}},\ \bibinfo {pages} {431} (\bibinfo {year}
  {2012})}\BibitemShut {NoStop}%
\bibitem [{\citenamefont {Wensink}\ \emph {et~al.}(2012)\citenamefont
  {Wensink}, \citenamefont {Dunkel}, \citenamefont {Heidenreich}, \citenamefont
  {Drescher}, \citenamefont {Goldstein}, \citenamefont {L{\"o}wen},\ and\
  \citenamefont {Yeomans}}]{Wensink:2012ci}%
  \BibitemOpen
  \bibfield  {author} {\bibinfo {author} {\bibfnamefont {H.~H.}\ \bibnamefont
  {Wensink}}, \bibinfo {author} {\bibfnamefont {J.}~\bibnamefont {Dunkel}},
  \bibinfo {author} {\bibfnamefont {S.}~\bibnamefont {Heidenreich}}, \bibinfo
  {author} {\bibfnamefont {K.}~\bibnamefont {Drescher}}, \bibinfo {author}
  {\bibfnamefont {R.~E.}\ \bibnamefont {Goldstein}}, \bibinfo {author}
  {\bibfnamefont {H.}~\bibnamefont {L{\"o}wen}}, \ and\ \bibinfo {author}
  {\bibfnamefont {J.~M.}\ \bibnamefont {Yeomans}},\ }\href@noop {} {\bibfield
  {journal} {\bibinfo  {journal} {Proceedings of the National Academy of
  Sciences of the United States of America}\ }\textbf {\bibinfo {volume}
  {109}},\ \bibinfo {pages} {14308} (\bibinfo {year} {2012})}\BibitemShut
  {NoStop}%
\bibitem [{\citenamefont {Giomi}\ \emph {et~al.}(2013)\citenamefont {Giomi},
  \citenamefont {Bowick}, \citenamefont {Ma},\ and\ \citenamefont
  {Marchetti}}]{Giomi:2013ky}%
  \BibitemOpen
  \bibfield  {author} {\bibinfo {author} {\bibfnamefont {L.}~\bibnamefont
  {Giomi}}, \bibinfo {author} {\bibfnamefont {M.~J.}\ \bibnamefont {Bowick}},
  \bibinfo {author} {\bibfnamefont {X.}~\bibnamefont {Ma}}, \ and\ \bibinfo
  {author} {\bibfnamefont {M.~C.}\ \bibnamefont {Marchetti}},\ }\href@noop {}
  {\bibfield  {journal} {\bibinfo  {journal} {PRL}\ }\textbf {\bibinfo {volume}
  {110}},\ \bibinfo {pages} {228101} (\bibinfo {year} {2013})}\BibitemShut
  {NoStop}%
\bibitem [{\citenamefont {Zhou}\ \emph {et~al.}(2014)\citenamefont {Zhou},
  \citenamefont {Sokolov}, \citenamefont {Lavrentovich},\ and\ \citenamefont
  {Aranson}}]{Zhou:2014gl}%
  \BibitemOpen
  \bibfield  {author} {\bibinfo {author} {\bibfnamefont {S.}~\bibnamefont
  {Zhou}}, \bibinfo {author} {\bibfnamefont {A.}~\bibnamefont {Sokolov}},
  \bibinfo {author} {\bibfnamefont {O.~D.}\ \bibnamefont {Lavrentovich}}, \
  and\ \bibinfo {author} {\bibfnamefont {I.~S.}\ \bibnamefont {Aranson}},\
  }\href@noop {} {\bibfield  {journal} {\bibinfo  {journal} {Proceedings of the
  National Academy of Sciences of the United States of America}\ }\textbf
  {\bibinfo {volume} {111}},\ \bibinfo {pages} {1265} (\bibinfo {year}
  {2014})}\BibitemShut {NoStop}%
\bibitem [{\citenamefont {Fily}\ \emph {et~al.}(2014)\citenamefont {Fily},
  \citenamefont {Baskaran},\ and\ \citenamefont {Hagan}}]{Fily:2014gy}%
  \BibitemOpen
  \bibfield  {author} {\bibinfo {author} {\bibfnamefont {Y.}~\bibnamefont
  {Fily}}, \bibinfo {author} {\bibfnamefont {A.}~\bibnamefont {Baskaran}}, \
  and\ \bibinfo {author} {\bibfnamefont {M.~F.}\ \bibnamefont {Hagan}},\
  }\href@noop {} {\bibfield  {journal} {\bibinfo  {journal} {Soft Matter}\
  }\textbf {\bibinfo {volume} {10}},\ \bibinfo {pages} {5609} (\bibinfo {year}
  {2014})}\BibitemShut {NoStop}%
\bibitem [{\citenamefont {Deseigne}\ \emph {et~al.}(2012)\citenamefont
  {Deseigne}, \citenamefont {L{\'e}onard}, \citenamefont {Dauchot},\ and\
  \citenamefont {Chat{\'e}}}]{Deseigne:2012kn}%
  \BibitemOpen
  \bibfield  {author} {\bibinfo {author} {\bibfnamefont {J.}~\bibnamefont
  {Deseigne}}, \bibinfo {author} {\bibfnamefont {S.}~\bibnamefont
  {L{\'e}onard}}, \bibinfo {author} {\bibfnamefont {O.}~\bibnamefont
  {Dauchot}}, \ and\ \bibinfo {author} {\bibfnamefont {H.}~\bibnamefont
  {Chat{\'e}}},\ }\href@noop {} {\bibfield  {journal} {\bibinfo  {journal}
  {Soft Matter}\ }\textbf {\bibinfo {volume} {8}},\ \bibinfo {pages} {5629}
  (\bibinfo {year} {2012})}\BibitemShut {NoStop}%
\bibitem [{\citenamefont {Woodhouse}\ and\ \citenamefont
  {Goldstein}(2012)}]{Woodhouse:2012vx}%
  \BibitemOpen
  \bibfield  {author} {\bibinfo {author} {\bibfnamefont {F.~G.}\ \bibnamefont
  {Woodhouse}}\ and\ \bibinfo {author} {\bibfnamefont {R.~E.}\ \bibnamefont
  {Goldstein}},\ }\href@noop {} {\bibfield  {journal} {\bibinfo  {journal}
  {arXiv}\ } (\bibinfo {year} {2012})},\ \Eprint
  {http://arxiv.org/abs/1207.5349v1} {1207.5349v1} \BibitemShut {NoStop}%
\bibitem [{\citenamefont {Bricard}\ \emph {et~al.}(2015)\citenamefont
  {Bricard}, \citenamefont {Caussin}, \citenamefont {Das}, \citenamefont
  {Savoie}, \citenamefont {Chikkadi}, \citenamefont {Shitara}, \citenamefont
  {Chepizhko}, \citenamefont {Peruani}, \citenamefont {Saintillan},\ and\
  \citenamefont {Bartolo}}]{Bricard:2015jx}%
  \BibitemOpen
  \bibfield  {author} {\bibinfo {author} {\bibfnamefont {A.}~\bibnamefont
  {Bricard}}, \bibinfo {author} {\bibfnamefont {J.-B.}\ \bibnamefont
  {Caussin}}, \bibinfo {author} {\bibfnamefont {D.}~\bibnamefont {Das}},
  \bibinfo {author} {\bibfnamefont {C.}~\bibnamefont {Savoie}}, \bibinfo
  {author} {\bibfnamefont {V.}~\bibnamefont {Chikkadi}}, \bibinfo {author}
  {\bibfnamefont {K.}~\bibnamefont {Shitara}}, \bibinfo {author} {\bibfnamefont
  {O.}~\bibnamefont {Chepizhko}}, \bibinfo {author} {\bibfnamefont
  {F.}~\bibnamefont {Peruani}}, \bibinfo {author} {\bibfnamefont
  {D.}~\bibnamefont {Saintillan}}, \ and\ \bibinfo {author} {\bibfnamefont
  {D.}~\bibnamefont {Bartolo}},\ }\href@noop {} {\bibfield  {journal} {\bibinfo
   {journal} {Nature Communications}\ }\textbf {\bibinfo {volume} {6}},\
  \bibinfo {pages} {1} (\bibinfo {year} {2015})}\BibitemShut {NoStop}%
\bibitem [{\citenamefont {Hiraoka}\ \emph {et~al.}(2017)\citenamefont
  {Hiraoka}, \citenamefont {Shimada},\ and\ \citenamefont
  {Ito}}]{Hiraoka:2017fw}%
  \BibitemOpen
  \bibfield  {author} {\bibinfo {author} {\bibfnamefont {T.}~\bibnamefont
  {Hiraoka}}, \bibinfo {author} {\bibfnamefont {T.}~\bibnamefont {Shimada}}, \
  and\ \bibinfo {author} {\bibfnamefont {N.}~\bibnamefont {Ito}},\ }\href@noop
  {} {\bibfield  {journal} {\bibinfo  {journal} {J. Phys.: Conf. Ser.}\
  }\textbf {\bibinfo {volume} {921}},\ \bibinfo {pages} {012006} (\bibinfo
  {year} {2017})}\BibitemShut {NoStop}%
\bibitem [{\citenamefont {Deseigne}\ \emph {et~al.}(2010)\citenamefont
  {Deseigne}, \citenamefont {Dauchot},\ and\ \citenamefont
  {Chat{\'e}}}]{Deseigne:2010gc}%
  \BibitemOpen
  \bibfield  {author} {\bibinfo {author} {\bibfnamefont {J.}~\bibnamefont
  {Deseigne}}, \bibinfo {author} {\bibfnamefont {O.}~\bibnamefont {Dauchot}}, \
  and\ \bibinfo {author} {\bibfnamefont {H.}~\bibnamefont {Chat{\'e}}},\
  }\href@noop {} {\bibfield  {journal} {\bibinfo  {journal} {Phys. Rev. Lett.}\
  }\textbf {\bibinfo {volume} {105}},\ \bibinfo {pages} {098001} (\bibinfo
  {year} {2010})}\BibitemShut {NoStop}%
\bibitem [{\citenamefont {Nguyen Thu~Lam}\ \emph {et~al.}(2015)\citenamefont
  {Nguyen Thu~Lam}, \citenamefont {Schindler},\ and\ \citenamefont
  {Dauchot}}]{Lam:2015bp}%
  \BibitemOpen
  \bibfield  {author} {\bibinfo {author} {\bibfnamefont {K.-D.}\ \bibnamefont
  {Nguyen Thu~Lam}}, \bibinfo {author} {\bibfnamefont {M.}~\bibnamefont
  {Schindler}}, \ and\ \bibinfo {author} {\bibfnamefont {O.}~\bibnamefont
  {Dauchot}},\ }\href@noop {} {\bibfield  {journal} {\bibinfo  {journal} {New
  Journal of Physics}\ }\textbf {\bibinfo {volume} {17}},\ \bibinfo {pages}
  {113056} (\bibinfo {year} {2015})}\BibitemShut {NoStop}%
\bibitem [{Note1()}]{Note1}%
  \BibitemOpen
  \bibinfo {note} {We used six different Cartesian coordinate systems in the
  six sectors of the hexagon, aligned with the respective boundaries. This
  allows to speak of parallel and perpendicular coordinates, $r_\parallel ,
  r_\perp $.}\BibitemShut {Stop}%
\bibitem [{\citenamefont {Briand}\ and\ \citenamefont
  {Dauchot}(2016)}]{Briand:2016fj}%
  \BibitemOpen
  \bibfield  {author} {\bibinfo {author} {\bibfnamefont {G.}~\bibnamefont
  {Briand}}\ and\ \bibinfo {author} {\bibfnamefont {O.}~\bibnamefont
  {Dauchot}},\ }\href@noop {} {\bibfield  {journal} {\bibinfo  {journal} {PRL}\
  }\textbf {\bibinfo {volume} {117}},\ \bibinfo {pages} {098004} (\bibinfo
  {year} {2016})}\BibitemShut {NoStop}%
\bibitem [{Note2()}]{Note2}%
  \BibitemOpen
  \bibinfo {note} {\protect \leavevmode {\protect Evidently, the
  presence of the boundaries dominates the global properties of flow and
  structure. This may be different for much larger systems, but this is pure
  speculation. We focus on finite systems, where the importance of boundaries
  even helps to keep the corresponding equilibrium system
  crystalline.}}\BibitemShut {Stop}%
\end{thebibliography}%

\end{document}